\documentclass[letterpaper,11pt]{article}

\usepackage[letterpaper,margin=1in]{geometry}

\usepackage{mathptmx}

\usepackage{tcolorbox}
\usepackage{fancyhdr}
\usepackage{float}
\usepackage{pdfpages}
\usepackage{relsize}
\usepackage{graphicx}
\usepackage{datetime2}

\newenvironment{titemize}{
\begin{itemize}
\setlength{\itemsep}{1pt}
\setlength{\parskip}{0pt}
\setlength{\parsep}{0pt}
}
{
\end{itemize}
}

\usepackage[hyphens]{xurl}
\usepackage{wrapfig}
\usepackage{hyperref}
\usepackage{graphicx}
\usepackage[numbers,sort&compress,square]{natbib}
\usepackage{doi}
\usepackage{color}
\definecolor{linkblue}{RGB}{0,0,180}
\hypersetup{
     colorlinks=true,
     citecolor=linkblue,
     filecolor=black,
     linkcolor=linkblue,
     urlcolor=linkblue
}

\usepackage{datetime}

\usepackage{titlesec}
\titlespacing{\paragraph} {0pt}{7pt}{5pt}

\begin{document}

\pagestyle{empty}
\includepdf[pages=1]{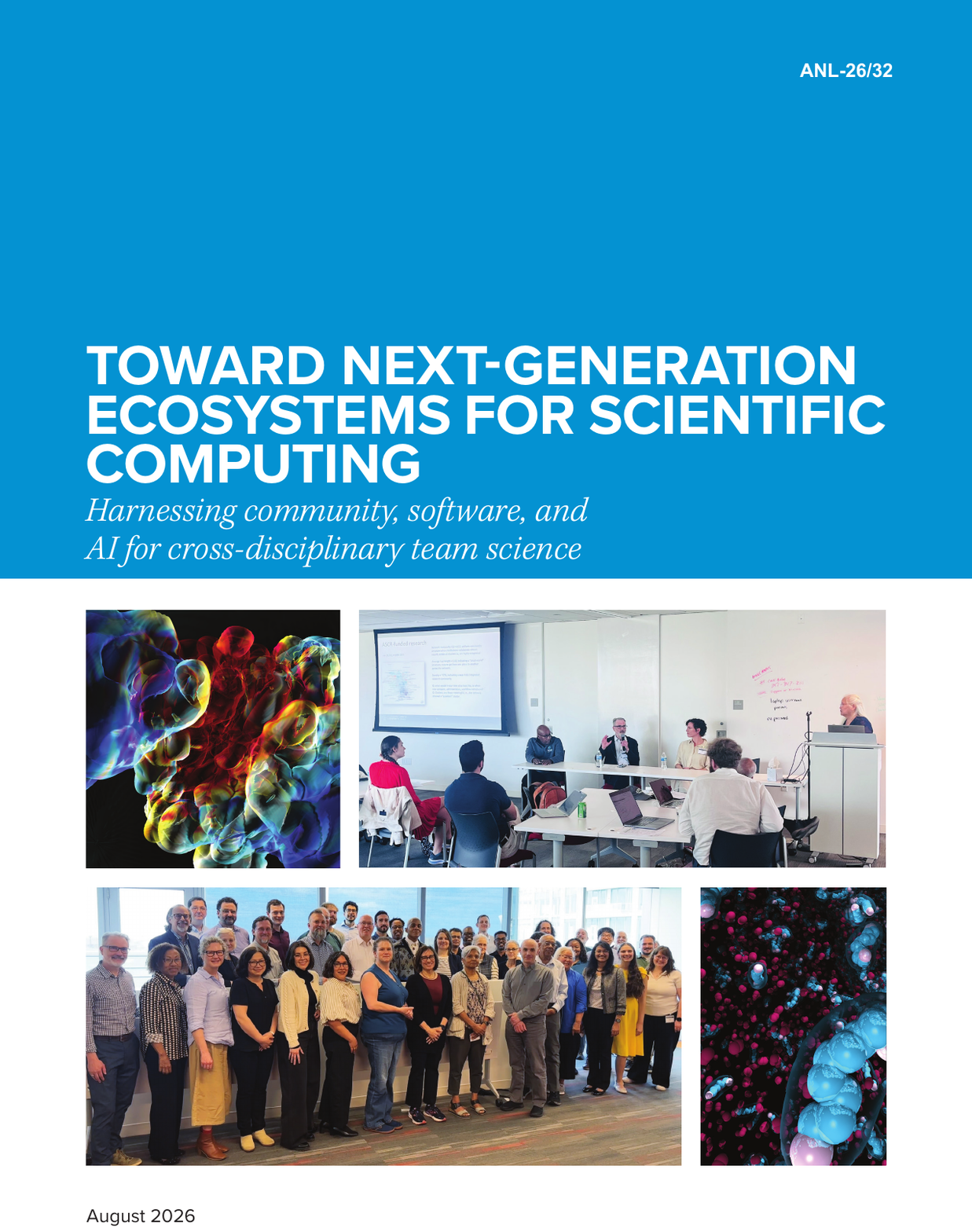}
\includepdf[pages=1]{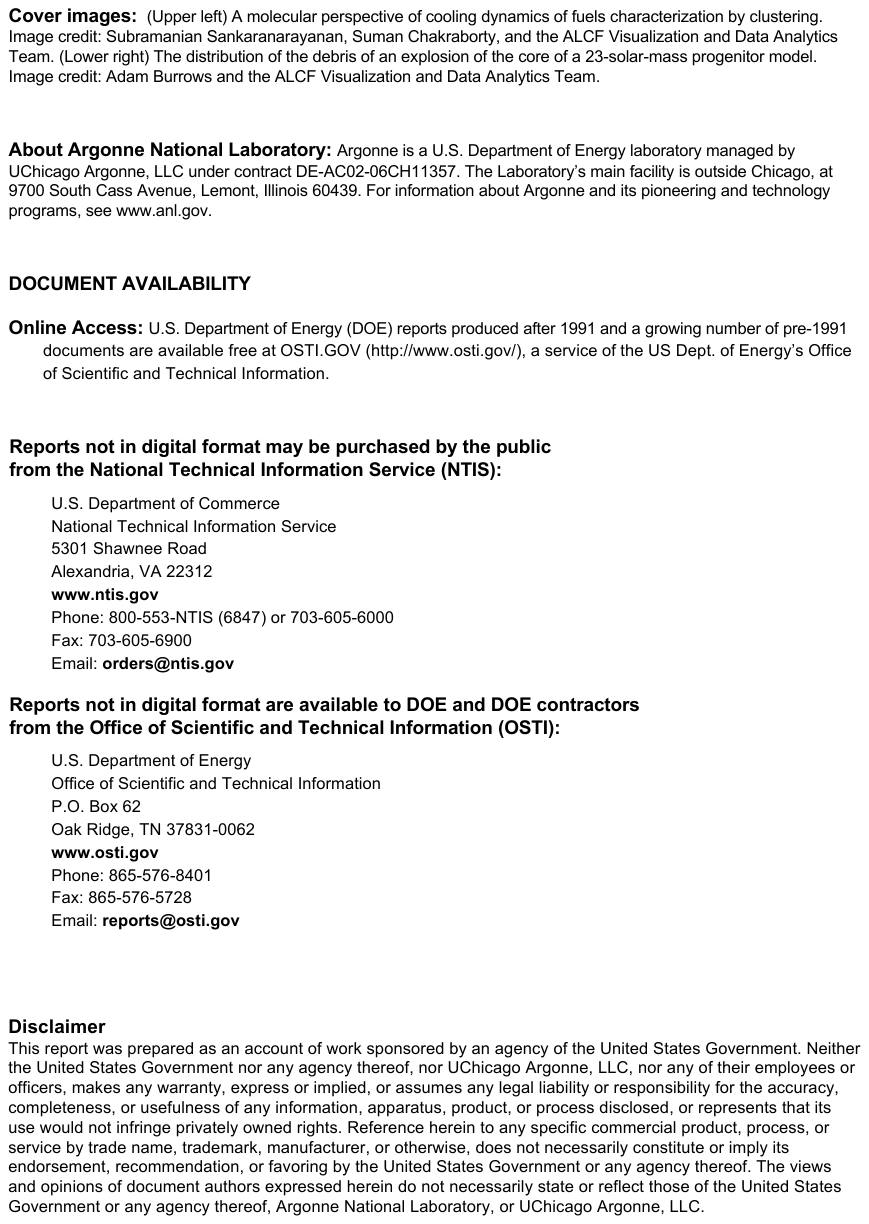}
\clearpage

\begin{center}
$\,$
\vspace{1.0in}

{\larger[2] \bf Report of the 2026 Workshop on}

\bigskip

{\larger[3] \bf Next-Generation Ecosystems for Scientific Computing:} \\

\smallskip

{\larger[1] \bf Harnessing Community, Software, and AI for Cross-Disciplinary Team Science}
\end{center}

\begin{center}


{\larger[1]

{\bf August 24, 2026}

\bigskip
\bigskip
{\bf Workshop Location:}\\
Chicago, IL

\bigskip
\bigskip

{\bf Workshop Dates:}\\
April 14--16, 2026
\bigskip
\bigskip
\bigskip
\bigskip
}

\end{center}

{\bf Suggested Citation}: Report of the 2026 Workshop on Next-Generation Ecosystems for Scientific Computing: Harnessing Community, Software, and AI for Cross-Disciplinary Team Science.
L. C. McInnes, D. Arnold, P. Balaprakash, M. Bernhardt, F. Cappello, B. Cerny, D. DiazGranados, A. Dubey, N. Etienne, R. Giles, D. G\'{o}mez-Zar\'{a}, D. W. Hood, M. A. Leung, V. L\'opez-Marrero, O. B. Newton, I. Qualters, K. Teranishi, S. M. Wild, G. Allen, R. Arthur, A. Ballow, T. Baylis, D. E. Bernholdt, D. Bielich, J. Cohoon, J. Crampton, C. Ferenbaugh, S. M. Fiore, T. Herault, T. Islam, S. Jacobsohn, M. Lin, C.~Lively, S. Matsuoka, S. Milojević, D. Nichols, C. Oehmen, S. Ospina Tabares, M. E. Papka, K. Riley, D.~Rouson, S. K. Seal, B. Segundo, J. Shalf, A. Siegel, V. Taylor, J. Willenbring, L. Woodley. Report ANL-26/32, 2026, 
\url{https://doi.org/10.48550/arXiv.2608.26519}.

\bigskip
\bigskip
\bigskip
\begin{center}
\vspace{-0.1in}
\includegraphics[width=1.0\textwidth] {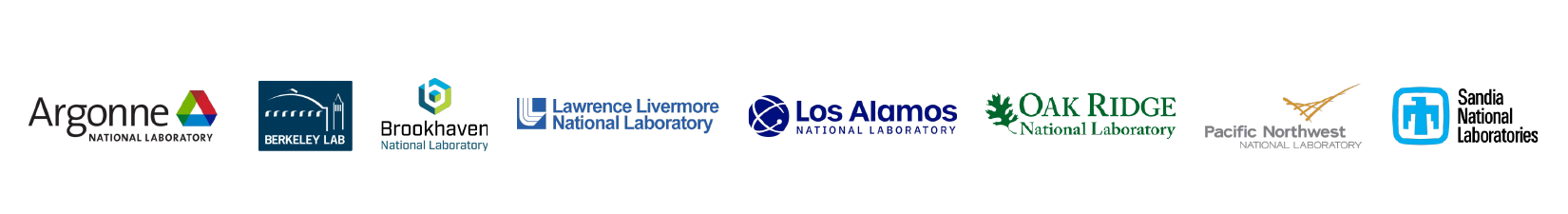}
\end{center}

\newpage

\begin{center}

{\large
{\bf Report Authors:}
}

{\small
\bigskip
Lois Curfman McInnes, Argonne National Laboratory\\
Dorian Arnold, Emory University\\
Prasanna Balaprakash, PrimaLabs\\
Mike Bernhardt, Team Libra\\
Franck Cappello, Argonne National Laboratory\\
Beth Cerny, Argonne National Laboratory\\
Deborah DiazGranados, Virginia Commonwealth University\\
Anshu Dubey, RIKEN Center for Computational Science\\
Nichole Etienne, Emory University\\
Roscoe Giles, Boston University\\
Diego G\'{o}mez-Zar\'{a}, University of Notre Dame\\
Denice Ward Hood, University of Illinois Urbana-Champaign\\
Mary Ann Leung, Sustainable Horizons Institute\\
Vanessa L\'opez-Marrero, Stony Brook University\\
Olivia B.\ Newton, University of Montana\\
Irene Qualters, Los Alamos National Laboratory (retired)\\
Keita Teranishi, Oak Ridge National Laboratory\\
Stefan M.\ Wild, Lawrence Berkeley National Laboratory\\

Gabrielle Allen, University of Wyoming\\
Richard Arthur, GE Aerospace\\
Alexandra Ballow, Montana State University\\
Tony Baylis, Lawrence Livermore National Laboratory\\
David E. Bernholdt, Oak Ridge National Laboratory\\
Daniel Bielich, Synopsys\\
Johanna Cohoon, Lawrence Berkeley National Laboratory\\
Jeremy Crampton, AAAS Fellow, U.S. Department of Energy\\
Charles Ferenbaugh, Los Alamos National Laboratory\\
Stephen M. Fiore, University of Central Florida\\
Thomas Herault, INRIA\\
Tanzima Islam, Texas State University\\
Stephen Jacobsohn, Google\\
Meifeng Lin, Brookhaven National Laboratory\\
Charles Lively, Lawrence Berkeley National Laboratory\\
Satoshi Matsuoka, RIKEN Center for Computational Science\\
Staša Milojević, Indiana University\\
Daniel Nichols, Lawrence Livermore National Laboratory\\
Chris Oehmen, Pacific Northwest National Laboratory\\
Santiago Ospina Tabares, University of Illinois Urbana-Champaign\\
Michael E. Papka, Argonne National Laboratory,  University of Illinois Chicago\\
Katherine Riley, Argonne National Laboratory\\
Damian Rouson, Lawrence Berkeley National Laboratory\\
Sudip K. Seal, Oak Ridge National Laboratory\\
Brittany Segundo, National Academies of Sciences, Engineering, and Medicine\\
John Shalf, Lawrence Berkeley National Laboratory\\
Andrew Siegel, Argonne National Laboratory\\
Valerie Taylor, Argonne National Laboratory\\
Jim Willenbring, Sandia National Laboratories\\
Lou Woodley, Center for Scientific Collaboration and Community Engagement\\
}
\end{center}
\clearpage

\pagestyle{plain}
\pagenumbering{roman}
\newpage
\tableofcontents

\newpage
\phantomsection
\section*{Executive Summary:\\ Toward Strategies for Trustworthy, AI-Enabled Scientific Computing}
\addcontentsline{toc}{section}{Executive Summary: Toward Strategies for Trustworthy, AI-Enabled Scientific Computing}

\bigskip

Scientific computing is entering a period of profound transformation. 
Advances in artificial intelligence, heterogeneous computing, emerging quantum capabilities, automation, and data-intensive research are reshaping not only computational tools but also the institutions, workforce models, and collaborative practices that underpin scientific discovery. 
The 2026 workshop {\em Toward Next-Generation Ecosystems for Scientific Computing} was the second in a three-year series and moved beyond identifying challenges to examining the actions required to address them. 
Building on the 2025 workshop, participants explored how to build scientific computing ecosystems that remain trustworthy, sustainable, innovative, and resilient in increasingly AI-enabled environments. 

The workshop identified four interdependent strategic  themes for next-generation scientific computing ecosystems: software ecosystems for AI-enabled discovery; trust, validation, and traceability; human-AI teaming and paradigm shifts; and workforce, pedagogy, and governance.
  
These themes are interdependent: weakness in any one can undermine the ecosystem as a whole.  
Powerful tools without adequate validation and traceability can produce results that are difficult to interpret, reproduce, or trust. 
Increased automation without effective human oversight can obscure assumptions, errors, and uncertainty, while technical advances unsupported by appropriate skills, governance, incentives, and collaborative practices may be misapplied or prove difficult to sustain. 
As AI systems assume larger roles in scientific workflows, future environments must enable people to understand, direct, evaluate, and intervene in computational processes. 
   
The report translates these themes into eight priorities  for community action aimed at helping ecosystems adapt to technological change while preserving scientific rigor and public trust.
Collectively, these workshop insights reinforce a central message: future scientific computing ecosystems must enable not only greater speed and efficiency but also deeper creativity, stronger collaboration, and more reliable pathways to discovery.

\clearpage
\pagenumbering{arabic}

\newpage

\section{Background and Motivation}

Scientific computing is at an ecosystem inflection point. 
For decades, high-performance computing has advanced science by enabling investigation of phenomena that cannot be studied through observation and experiment alone due to their scale, complexity, cost, or risks \cite{hendrickson2020ascr,GroppHarrisonEtAl2016,KeyesTaylor2011,siam-cse18,NSF-OAC-Blueprint2021,siam-futurecse2025}. 
Computational methods help
researchers understand, predict, and shape outcomes across domains including energy, materials, health, engineering, security, and fundamental discovery.
Today, AI, heterogeneous architectures, specialized accelerators, and distributed data resources 
are reshaping the tools and environments that support discovery.
Because scientific computing underpins industrial competitiveness, engineering design, national security, and public decision-making, the challenge is not simply to build faster machines or more powerful tools. It is to create ecosystems that can incorporate new technologies without compromising the scientific rigor, trust, and human judgment on which their value ultimately depends.
Without coordinated development, these advances may amplify fragmentation, technical debt, misplaced trust, uneven access, and accountability challenges.

{\bf About this report.} 
This report summarizes the 2026 Workshop on {\em Next-Generation Ecosystems for Scientific Computing: Harnessing Community, Software, and AI for Cross-Disciplinary Team Science}, held April 14--16, 2026, in Chicago, IL.\footnote{Appendices \ref{sec:workshop-description}, \ref{sec:workshop-participants}, and \ref{sec:workshop-agenda} provide the workshop description, list of participants, and agenda, respectively.} 
The workshop brought together experts from high-performance computing (HPC), AI, cognitive and social sciences, community development, and related fields to identify priorities for building trustworthy, adaptive, and sustainable ecosystems for discovery.
This report is intended for research communities, funders, laboratories, universities, private-sector partners, and professional societies 
working to strengthen these ecosystems.

{\bf Workshop series, objectives, and progression.}
This report is part of a three-year workshop series 
that uses a socio-technical co-design perspective to examine how technical innovation, team science, workforce development, governance, and community stewardship must evolve together for next-generation scientific computing.  
The first workshop, held in 2025, established the foundation for the series by framing scientific computing ecosystems as dynamic  systems in which hardware, software, AI, human expertise, institutions, and community practices coevolve~\cite{Workshop-next-generation-ecosystems2025,ecosystems-cise}. 
It identified three connected challenge areas—AI-integrated software ecosystems, cross-disciplinary collaboration, and pedagogy and workforce development—and emphasized the need for modular and trustworthy software, effective integration of AI into scientific teams while preserving creativity and rigor, and adaptive training pathways.

The 2026 workshop builds directly on this foundation while advancing the series from identifying challenges toward developing strategic priorities. 
It asks what capabilities, norms, and coordinated actions are now required as AI becomes a more active participant in scientific work. 
This synthesis refines the earlier challenge areas into four interdependent strategic themes, elevates trust and governance as explicit ecosystem requirements, and translates the discussion into eight priorities for community action. 
This progression is important because the recurring concerns involving software, collaboration, workforce development, and sustainability are persistent structural issues, while rapid advances in agentic AI are increasing their urgency and changing how they must be addressed.

{\bf Scientific computing ecosystems as socio-technical systems.}
Socio-technical systems thinking emphasizes that technical capabilities and the human and organizational structures through which they are developed and used must be designed together rather than optimized separately~\cite{Trist1981SociotechnicalSystems}.
Scientific computing has long reflected this principle through the co-evolution of hardware, software, numerical methods, domain expertise, and collaborative practices. 
As scientific questions have become more complex, progress has increasingly depended on interconnected ecosystems that coordinate technology, expertise, and organizational structures across a diverse community of stakeholders, from academia and national laboratories to the private sector, government, and international collaborators.
This distributed structure underscores the importance of team science~\cite{nas-teams2015,nas-teams2025} and of
integrating AI not only into research methods but also into the software ecosystems that enable, connect, and sustain scientific work~\cite{DOE-WorkshopReportAI4Science2023,sssdu-workshop-report2023,NAIRR-pilot,nairr-software-workshop-report2025,nas-nnsa2023,AI-for-science2020,LaxRetrospective2026}.

In this report, {\em ecosystem} refers to interconnected technical (hardware, software, data, models, workflows), social (people, community practices, incentives), and organizational (institutions, governance) structures that shape scientific computing. 
The goal is not to design a static arrangement of components, 
but to identify conditions under which these ecosystems can sustain themselves, respond to change, and improve over time.
Progress ultimately depends on people with the expertise to shape and responsibly apply emerging technologies~\cite{ascac-workforce2014, ASCAC_ReportGiles2020,web-supercharging-americas-ai-workforce,siam-futurecse2025,ASCACCSGF2025,AIAA-WD1,LaxRetrospective2026}, institutions that foster collaboration and stewardship, and sustained contributions from government, academia, national laboratories, nonprofits, and the private sector.
The private sector is especially important both as a major user of scientific computing and as a source of the technologies and services that support modern workflows.

{\bf From AI tools to agentic scientific ecosystems.}
The most consequential change since 2025 is the shift from AI as a tool used within scientific workflows toward AI as an active participant in them.
AI systems now generate hypotheses and code, synthesize information, propose workflow steps, discover and optimize algorithms, coordinate tasks, assist with interpretation, and support surrogate model construction. 
This shift from a ``tool era'' toward an ``agentic workflow era'' does not reduce the need for human expertise; it changes where that expertise is most needed~\cite{prasanna-presentation2026.04,denario_project,novikov2025alphaevolve,Ghareeb_2026,Aygun_2026,10.1145/3769314,vriza_2026,shao_2026_SciSciGPt,Lu_2026_AI_Scientist,10.1145/3809164}. 
As automated and semi-autonomous systems become more deeply embedded in scientific work, longstanding
concerns about provenance, validation, attribution, and accountability become more urgent.
Traditional mechanisms for evaluating and governing scientific work may be incomplete when contributions are distributed across people and technological systems.  
Communities thus need records of the decisions, assumptions, and computational processes that shape results, along with standards specifying which artifacts are required for review, validation, and accountability.
 
Against this backdrop, the 2026 workshop focused on how to preserve and improve scientific quality as AI assumes larger roles in research.  
The objective is not simply faster science, but better science: research that produces valid and useful insights, makes assumptions and uncertainty visible, supports scrutiny and reproducibility where possible, and expands the questions and approaches scientists can pursue. 
Although AI can accelerate individual tasks, scientific progress still depends on judgment-intensive work such as framing questions, integrating expertise, selecting appropriate methods, interpreting results, managing uncertainty, and deciding when evidence is sufficiently trustworthy.
Accordingly, the report examines how technical capabilities, user experience, team norms, workforce development, governance, and incentives can be aligned so that AI strengthens the quality, creativity, and reliability of discovery rather than merely increasing output.  
Progress toward this goal is under way, but important questions remain~\cite{zhao_2025,10.1145/3773295,shapira2026agentschaos,10.1145/3722476,10.1145/3747200,10.1145/3749447,10.1145/3748642,10.1145/3743165,10.1145/3744911,10.1145/3789199,Nama_2026}.

{\bf Related community efforts.} 
Several recent reports and activities are complementary to this workshop series.  
The NSF workshop report on AI for the mathematical and physical sciences \cite{Ferguson_2026} examines how those communities and AI can mutually strengthen each other, while 
the Society for Industrial and Applied Mathematics (SIAM) Task Force Report \cite{SIAM_AI_Task_Force_Report_2026} emphasizes applied mathematics in developing trustworthy and interpretable AI for predictive science. 
The National Academies of Sciences, Engineering, and Medicine (NASEM) report on foundation models for scientific discovery \cite{nasem-foundation-models2025} considers how such models can complement traditional computational methods and recommends strategies for DOE-relevant scientific discovery.
Related NASEM studies address team science~\cite{nas-teams2025}, human-AI teaming \cite{NASEM_human-ai_teaming}, 
and human and organizational factors in AI risk management \cite{nasem-human-ai2025}. 
Workshops organized by the Computing Research Association~\cite{ccc-workshop2026} and Research Software Alliance~\cite{resa-workshop2026,resa-workshop2-outcomes026} address trustworthy AI-enabled software systems and the changing role of research software engineering.

Across these efforts, recurring needs include standards, best practices, and evaluation frameworks for trustworthy AI-enabled research. Current work addresses 
the responsible use of agentic AI~\cite{CISA_2026}, identification and authorization of AI agents~\cite{NIST_2026}, and evaluation of AI scientific assistants~\cite{eaira_framework}. 
Additional work addresses evolving 
publication and peer-review practices~\cite{Kolda_2026}, detection of hallucinated references~\cite{refchecker}, and AI-assisted coding, software optimization, and  research support \cite{Garfinkel_2026,10.1145/3773295, nichols_2025,denario_project,riosgarcia_2026}. 
Together, these efforts reinforce the need to develop technical capabilities, evaluation practices, and community norms in concert.
The workshop synthesized these concerns into four interdependent strategic themes introduced in Section~\ref{sec:themes}.

{\bf Workshop approach and synthesis process.} 
As shown in Appendix~\ref{sec:workshop-agenda}, the workshop combined invited presentations, panel discussions, facilitated breakout sessions, and plenary report-outs. 
Breakout groups examined technical, organizational, workforce, and governance questions from complementary perspectives. 
The writing team reviewed session notes and report-outs to identify recurring concerns, proposed actions, and areas requiring further study. 
These findings were consolidated into four strategic themes and eight priority areas and refined through multiple rounds of participant review. 
The resulting report represents a qualitative synthesis of workshop discussions rather than a formal consensus process or systematic literature review.
The four themes are discussed in Section~\ref{sec:themes}, and the eight priorities for community action are presented in Section~\ref{sec:paths-forward}.

\section{Strategic Themes}
\label{sec:themes}

\begin{figure}[bht]
\centering
\vspace{-0.1in}
\includegraphics[width=0.85\textwidth]{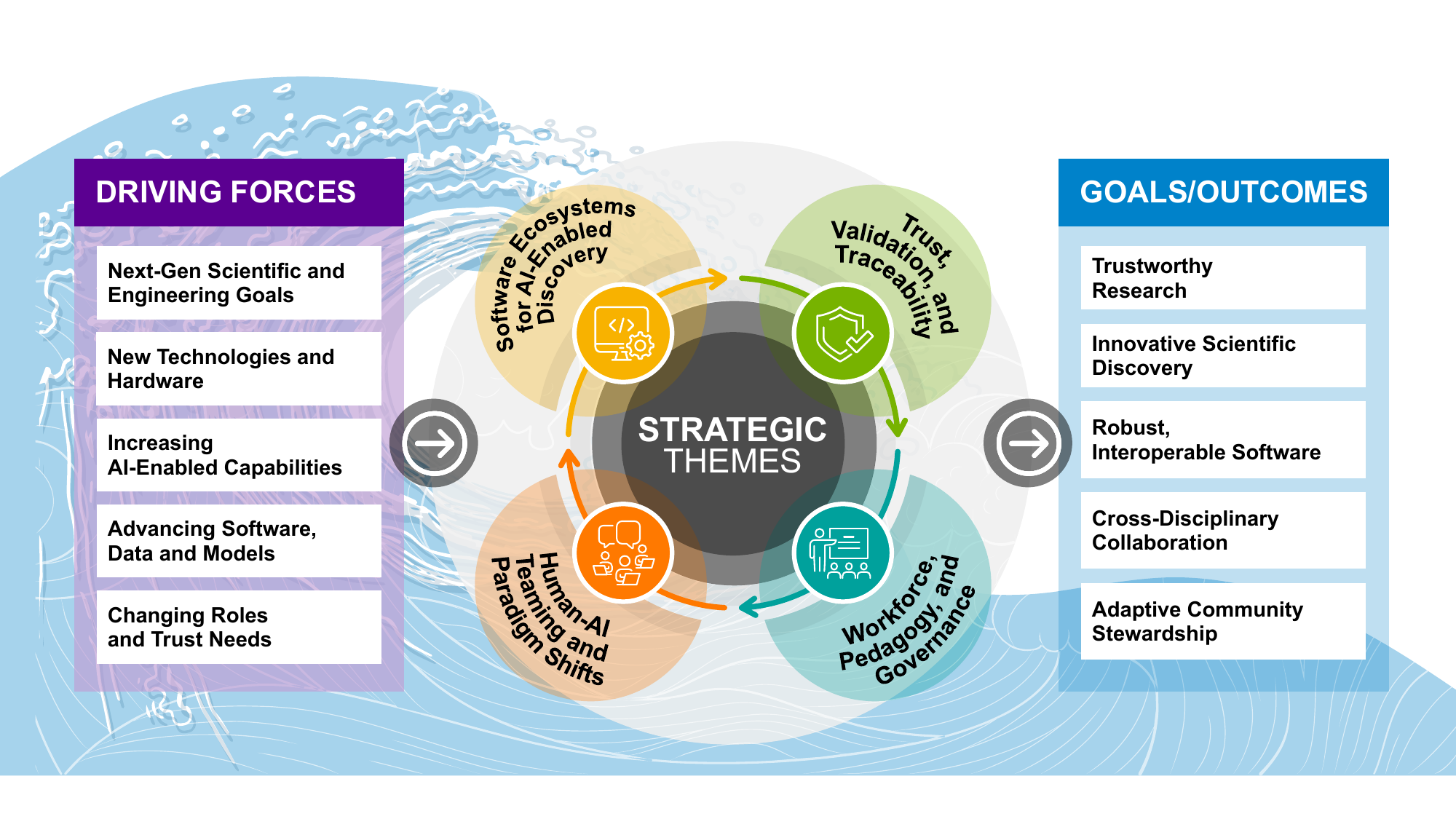}
\vspace{-0.2in}
\caption{
Strategic themes for next-generation scientific computing ecosystems.
The wave represents forces creating urgency, while the central cycle links four interdependent themes to desired outcomes.
}
\label{fig:workshop-themes}
\end{figure}

The 2026 workshop discussions converged on four interdependent strategic themes that cut across the technical, social, institutional, and educational dimensions of scientific computing. As Figure~\ref{fig:workshop-themes} illustrates, these themes
link broad drivers of change to desired outcomes for next-generation ecosystems.    
They should be understood not as separate components, but as interacting design concerns that shape how ecosystems learn, adapt, and remain sustainable. 
This section explains the capabilities associated with each theme;
Section~\ref{sec:paths-forward} develops eight related priority areas for community action. Some align primarily with one theme, while others draw on several.

{\addcontentsline{toc}{subsection}{Software ecosystems for AI-enabled scientific discovery}
\paragraph{Software ecosystems for AI-enabled scientific discovery.}
Next-generation scientific discovery depends on trustworthy and usable software ecosystems that are also interoperable, portable, and sustainable across rapidly changing computing environments.
These ecosystems must support both AI-enabled components and the layered software stacks required by modern scientific 
applications.\footnote{For example, see the wide range of scientific applications within the recent DOE Exascale Computing Project (ECP)~\cite{ecp-website,alexander_exascale_2020}.} 
Scientific codes build on low-level programming models and runtimes, mathematical and visualization libraries, performance tools, emerging ML/AI technologies, and  application-specific components.\footnote{For example, E4S~\cite{E4S-web} provides a foundational HPC-AI software ecosystem for science~\cite{HerouxEtAl2024}, including ECP libraries and tools~\cite{SWEcosystems:NCS2021,ECP-software-technologies2024,TransformingScienceThroughSoftware2024} as well as popular AI packages. The Consortium for the Advancement of Scientific Software (CASS)~\cite{cass-website} and its member organizations coordinate the stewardship of E4S and many of its constituent packages.}

Tool discoverability and sustainment remain essential, as do validation and interoperability.
A recurring theme was the importance of shared research assets that capture knowledge, context, and operational experience in forms that can be reused across projects and communities. 
These resources are not merely supporting materials; they are essential infrastructure that enables distributed teams to understand, extend, validate, and ultimately trust complex scientific computing workflows.
Digital-twin and digital-thread frameworks 
illustrate how models, data, software, and lifecycle context connect through traceable, reusable representations~\cite{AIAA-DT1,AIAA-DT2,AIAA-Th1}.
Shared research assets are especially important as workflows combine conventional simulation, AI-enabled components, accelerator-based computing, and emerging quantum approaches, each of which may introduce assumptions, performance constraints, validation requirements, and reproducibility challenges.
 
As AI-enabled workflows become more common, scientific computing ecosystems should help researchers select computational approaches that are appropriate for the problem at hand rather than defaulting to the most powerful or resource-intensive available option. Different scientific tasks require different balances of accuracy, interpretability, reproducibility, performance, and cost, and in many cases simpler or more specialized approaches may provide the best solution \cite{10.1145/3771728,bernadotte_2026,duggan2026,spaan2026reducingcomputewastellms}. Effective ecosystem design should therefore provide the guidance needed to align computational methods with scientific objectives, validation requirements, and available resources, promoting both scientific rigor and responsible use of computing infrastructure.

Beyond the software stack itself, user experience emerged as a core ecosystem concern.
In this context, user experience extends far beyond interfaces to encompass the entire process by which researchers discover, learn, access, and effectively use computational capabilities. The quality of that experience strongly influences who can participate, which tools are adopted, and whether those tools are applied appropriately. Poorly designed experiences can create barriers for new users and widen gaps between experts and non-experts. Lack of sufficient guidance can lead to the use of methods in ways that produce plausible but scientifically invalid results.

Participants emphasized that effective ecosystem design is not synonymous with eliminating friction. Rather, it involves directing effort and attention where they are most needed. Routine, well-understood activities should be straightforward, while actions that carry greater uncertainty, risk, cost, or scientific consequence should provide opportunities for additional feedback, scrutiny, and informed decision-making.

The discussions also pointed to researcher-facing support systems that begin with the scientist’s problem and help identify relevant data, available tools, missing pieces, constraints, and next steps. 
Rather than simply generating outputs, these systems should help users navigate the broader ecosystem. 
This approach remains valuable even as specific AI technologies evolve because it addresses a more fundamental need: connecting scientific problems with the methods, resources, expertise, and evidence required to advance discovery responsibly and sustainably.
Usability alone, however, is not sufficient: researchers must also be able to trace, evaluate, and challenge the results these systems produce.

\paragraph{Trust, validation, and traceability.}
{\addcontentsline{toc}{subsection}{Trust, validation, and traceability} 
Trust in next-generation scientific computing ecosystems requires that  
verification, validation, and uncertainty quantification (VVUQ), 
provenance, traceability, transparency, and auditability become core design requirements rather than afterthoughts. 
In increasingly agentic workflows, these are also essential to system-level evaluation and safety because they determine whether AI-enabled results can be interpreted, challenged, reused, and trusted.
The 2026 discussions emphasized trust frameworks that span the full workflow.  
The NIST AI Risk Management Framework~\cite{NIST-ai-risk-management-framework-2026} provides additional context for operationalizing risk management, evaluation, and governance in AI-enabled systems.
These frameworks must capture the full context needed to understand, evaluate, and reproduce computational results. Alongside traditional information about software, data, and execution environments, they must also capture the decisions, assumptions, intermediate processes, and human contributions that increasingly shape modern computational workflows.

A recurring concern was that existing VVUQ methods may be inadequate
as AI-enabled workflows move beyond exploratory use into scientific interpretation, design, policy analysis, and operational decision-making.
These methods must be extended and integrated with approaches tailored to hybrid workflows that combine mechanistic models, numerical solvers, surrogate models, data-driven components, agentic systems, and human judgment. 
In such settings, uncertainty assessment must span the disciplines involved in the workflow  
to evaluate confidence, limitations, regimes of validity, and implications for decision-making~\cite{Arthur2026DecisionMaking}.

Scientific software also has trust requirements that differ from many other forms of software.
It can execute without error yet still produce misleading results when numerical methods are inappropriate, convergence is inadequate, or outputs violate physical or domain constraints. 
Conventional test suites remain necessary, but they may not detect these failures, especially when AI generates or modifies software faster than experts can inspect it.
New methodologies are therefore needed, including mathematically and statistically grounded diagnostics and, where feasible, formal reasoning.
When exact reproducibility is impossible because hardware, software, or operational conditions change, traceability and uncertainty quantification become essential for judging whether results remain credible.
Auditability makes these practices visible, reviewable, and actionable across the ecosystem. 
Such capabilities become even more important when AI systems participate actively in workflows rather than support isolated tasks.
 
\addcontentsline{toc}{subsection}{Human-AI teaming and paradigm shifts}
\paragraph{Human-AI teaming and paradigm shifts.}
Building on the 2025 workshop’s emphasis on cross-disciplinary collaboration and AI for scientific software teams, participants reinforced a central theme: as AI takes on more taskwork, teamwork becomes even more important. 
Research on team effectiveness distinguishes {\em taskwork} (the activities required to accomplish team goals) from {\em teamwork} (how members work together effectively toward those goals)~\cite{competencies1995}; 
both shape performance in collaborative science~\cite{fiore2008}, and
this distinction remains relevant for human-AI teaming~\cite{Bendell2025ArtificialSocialIntelligence}.
AI systems are becoming increasingly capable of scientific taskwork, but proficiency at individual tasks does not by itself ensure effective teaming. 
The central question is how humans and AI can combine their capabilities to produce reliable, interpretable, and scientifically meaningful results.

A key design choice is whether AI-enabled systems emphasize automation or augmentation. 
Automation can increase speed and efficiency, but it can weaken scientific understanding when it obscures assumptions, intermediate steps, uncertainty, or failure modes. 
Augmentation extends researchers' capabilities while preserving their ability to understand how results were produced, judge whether those results are plausible, and intervene when needed. 
The appropriate balance will vary by task, but activities involving greater uncertainty, risk, or scientific consequence require correspondingly greater transparency, validation, and human judgment.

Agentic and multi-agent systems intensify these coordination challenges~\cite{Guo2024LLMMultiAgents,Ghafarollahi2025SciAgents}. 
Adding agents does not necessarily improve a workflow: they may duplicate effort, rely on incompatible assumptions, propagate errors, or fail to recognize when additional expertise is needed. 
Scientific computing ecosystems therefore need mechanisms that maintain shared goals and task states, expose assumptions and limitations, preserve provenance, reconcile conflicting outputs, and escalate questions when uncertainty or required authority exceeds an agent's scope.

Human researchers remain central not only for oversight but also for framing scientific questions, integrating domain and computational knowledge, evaluating evidence, managing uncertainty, and interpreting unexpected results. 
As intelligent systems assume more computational work, people may increasingly direct, evaluate, and integrate AI-generated contributions within broader scientific objectives. 
Organizations will therefore need explicit norms for where automation is appropriate, when human review is required, how responsibility is assigned, and how decisions are made when evidence is incomplete. 
Human-AI teams should be evaluated not only by speed or output but also by scientific validity, error detection, reproducibility, recovery from failure, quality of collaboration, and whether researchers retain sufficient understanding to explain and extend the work. 
Shared expectations will also help align workflows as institutions adopt AI-enabled practices at different rates.

These changing roles reinforce the need to learn cross-disciplinary collaboration and human-AI teaming through experience.
The 2026 workshop emphasized exposing early-career researchers to authentic teaming environments through graduate fellowships, laboratory internships, industry apprenticeships, and cross-sector research projects. 
Public-private partnerships built around iterative co-design, co-development, and co-delivery provide another pathway~\cite{LeungEtAl-RFI-2026}.
Examples include the U.S.\ Department of Energy's (DOE's) 
Exascale Computing Project~(ECP; \cite{ecp-website,ecp-kothe-lee-qualters-2019}), 
Scientific Discovery through Advanced Computing (SciDAC) program~\cite{scidac-website}, and 
Computational Science Graduate Fellowship (CSGF) program~\cite{CSGF2021, csgf-website};
the National Nuclear Security Administration's (NNSA's) 
Predictive Science Academic Alliance Program~\cite{nnsa-psaap-website}; the
National Science Foundation's (NSF's) 
Science and Technology Centers~\cite{nsf-stc-website} and 
Engineering Research Centers~\cite{nsf-erc-website}; and
Sustainable Research Pathways~\cite{SRP-website}. 
These initiatives combine technical education with hands-on, cross-sector experience at scale. 
Comparable international and cross-border programs should also inform preparation for globally distributed scientific computing ecosystems.

These experiences help early-career researchers develop the boundary-spanning skills needed to work across disciplines while demonstrating how team-based computing contributes to discovery, engineering impact, and societal benefit. 
Rewarding career paths for people who combine technical depth with collaborative leadership therefore connect the teaming and workforce agendas.

\addcontentsline{toc}{subsection}{Workforce, pedagogy, and governance}
\paragraph{Workforce, pedagogy, and governance.}
Workforce development must move beyond technical proficiency to prepare people
to exercise judgment within complex socio-technical ecosystems. 
Roles in HPC software teams are also changing: research software engineers are becoming more professionalized, while AI-enabled tools are making it possible for more
non-specialists to participate in software creation and adaptation. 
Broader discussions, including the NASEM report on AI and the future of work~\cite{nasem-ai-work2025}, reinforce the need to prepare workers and institutions to adapt flexibly as AI changes tasks, roles, and skill requirements.  
These shifts create both opportunity and risk: more people can contribute, but many will need stronger preparation to validate, maintain, and govern what is produced.

Because workforce preparation emerged as a concrete implementation need, Figure~\ref{fig:workforce-roadmap} presents a roadmap linking foundational priorities, knowledge topics, competencies, delivery pathways, and intended outcomes. 
Training must extend beyond coding to prepare individuals for the broader intellectual demands of modern scientific computing. This includes developing a common foundation in evaluating evidence, understanding uncertainty, and assessing the reliability of computational results. Applied mathematics remains essential to this effort, providing many of the core concepts that support scientific modeling, computation, and decision-making.

\begin{figure}[bht]
\centering
\includegraphics[width=0.85\textwidth]{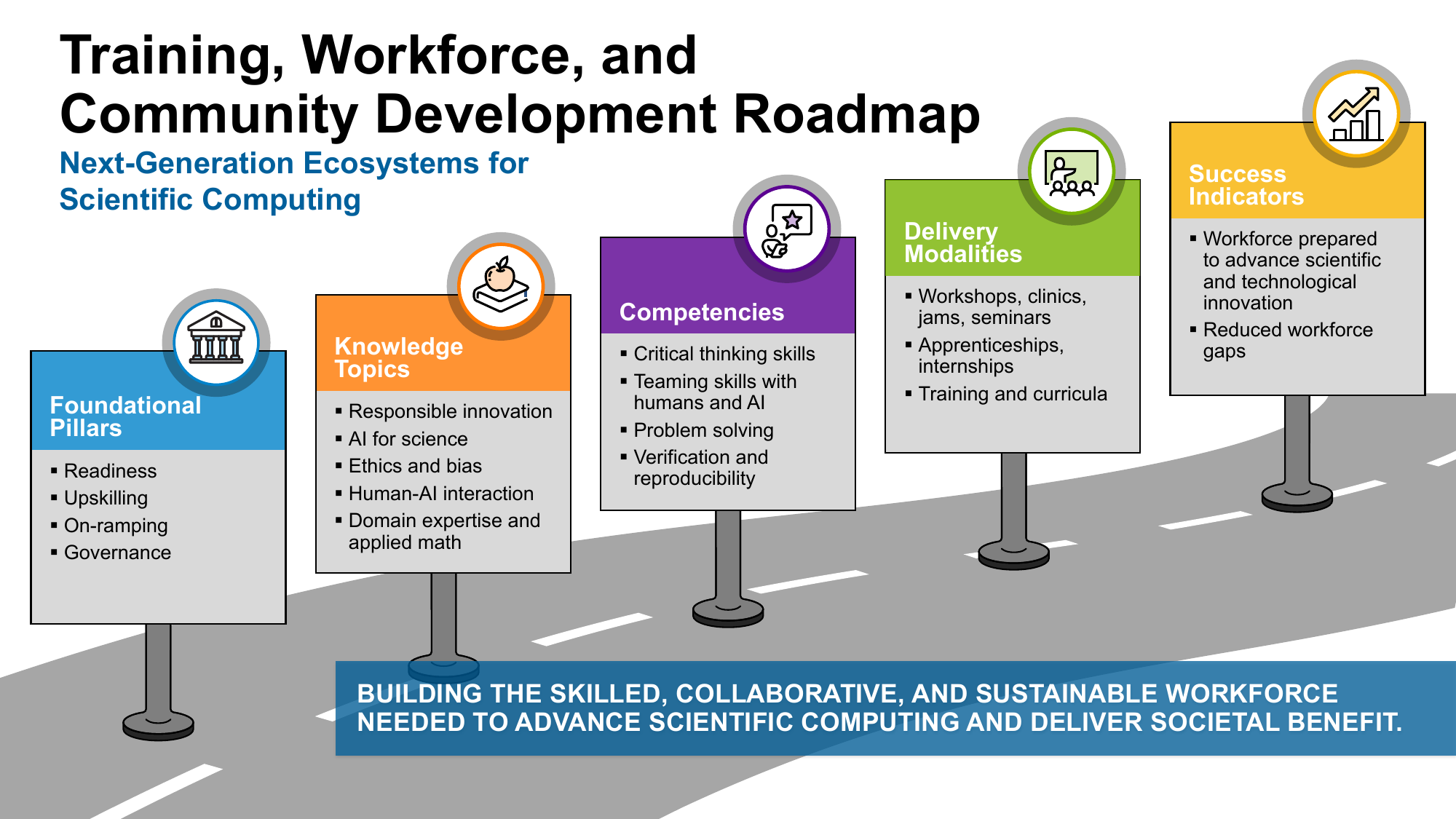}
\caption{
Roadmap for training, workforce, and community development. Foundational pillars, knowledge topics, competencies, and delivery pathways jointly support a skilled, collaborative, and sustainable workforce.}
\label{fig:workforce-roadmap}
\end{figure}

A resilient workforce strategy should provide multiple pathways for learning and professional development, recognizing that people enter scientific computing at different career stages and in different roles. 
Effective programs should support newcomers entering the field, help practitioners adapt to evolving technologies and practices, and prepare leaders to guide the governance and stewardship of increasingly complex ecosystems. 
The objective is not to train people for a single generation of tools, but to cultivate competencies that remain valuable as technologies evolve. 
These include durable taskwork competencies (e.g., problem solving) and teamwork competencies (e.g., communication) that support effective collaboration across settings~\cite{competencies1995}.
Success should be measured by whether the ecosystem develops a workforce capable of advancing scientific and technological innovation while reducing persistent workforce gaps.
Workforce strategy should also distinguish short-, medium-, and long-term needs. 
Near-term training can emphasize durable competencies and responsible use of tools; medium-term efforts can build experience with human-AI teaming and evolving workflows; and longer-term work should study how AI changes scientific roles, behavior, institutions, and career pathways.

\section{Paths Forward: Community Actions}
\label{sec:paths-forward}

The four themes above describe the capabilities needed in next-generation scientific computing ecosystems. 
Because no single project or organization can provide the necessary infrastructure, norms, incentives, and workforce pathways, progress will require coordinated community action.
The following eight priorities offer complementary directions for research communities, institutions, and funders. 
Some build on established practices; others require pilots and continued study.
\begin{titemize}
\item
\addcontentsline{toc}{subsection}{Build and sustain shared research assets and open infrastructure}
{\bf Build and sustain shared research assets and open infrastructure.} 
Treating curated libraries, datasets, metadata standards, reusable computational pipelines, benchmarks, validation examples, and documentation as first-class infrastructure can strengthen reuse, continuity, and long-term sustainability. 
Maintaining and improving these assets—and appropriately valuing those who steward them—are as important as creating them.

\item
\addcontentsline{toc}{subsection}{Build capabilities for trust and traceability}
{\bf Build capabilities for trust and traceability.}
Trustworthy scientific processes increasingly depend on 
incorporating VVUQ, provenance, and auditability 
from the outset. 
Having a full trace of 
assumptions, intermediate decisions, human interventions, and execution context is important for 
reviewing, reproducing results where possible, and assessing when exact reproduction is not feasible.
As agentic systems become 
more deeply embedded in scientific processes, trust infrastructure must also contend with identity management, authorization, data governance, security monitoring, and mechanisms for containment and recovery.

\item
\addcontentsline{toc}{subsection}{Treat user experience as scientific infrastructure}
{\bf Treat user experience as scientific infrastructure.}
User experience influences whether researchers can identify methods, understand constraints, interpret outputs, and recognize when further validation is needed. 
Interfaces, access mechanisms, support structures, and policies can help users make informed choices throughout  research workflows. 
Effective design can lower unnecessary barriers while preserving scrutiny when uncertainty, cost, or consequences are significant.

\item
\addcontentsline{toc}{subsection}{Establish explicit norms for human-AI teaming}
{\bf Establish explicit norms for human-AI teaming.}
Human-AI teaming benefits from clear expectations for automation, human judgment, oversight, documentation, and accountability. 
These norms become actionable when organizations specify who may delegate work to an agent, when a result must be escalated for expert review, who can approve an AI-generated result for use, and how responsibility is shared across institutions.
Boundary-spanning roles remain important for connecting expertise and reducing misalignment across disciplines and organizations.

\item
\addcontentsline{toc}{subsection}{Center workforce development on judgment}
{\bf Center workforce development on judgment.}
Workforce preparation increasingly requires opportunities for students, practitioners, and institutional leaders to develop skills in collaboration, verification, reasoning under uncertainty, and responsible tool use. 
Technical training alone will not be sufficient for work involving AI-enabled systems, multiple forms of expertise, and distributed responsibility. 
Pilots and longitudinal studies can clarify how AI changes professional roles, team behavior, job satisfaction, and career pathways.

\item
\addcontentsline{toc}{subsection}{Coordinate across sectors and borders}
{\bf Coordinate across sectors and borders.} 
Because scientific computing ecosystems span academia, national laboratories, the private sector, government, nonprofits, and international partners, progress  depends on sustained coordination across these communities. 
Shared mechanisms can align standards, 
sustain common infrastructure, exchange lessons learned, reduce duplicated effort, and connect rapidly evolving technologies with long-term scientific needs.

\item
\addcontentsline{toc}{subsection}{Align incentives with stewardship and sustainability}
{\bf Align incentives with stewardship and sustainability.}
Institutions and funders should recognize and reward researchers who maintain scientific software, validate shared resources, support user communities, and preserve knowledge. 
Although often less visible than launching a new tool or publishing a new result, this work enables researchers to build on trusted foundations rather than repeatedly reconstructing them.
Without sustained funding and professional recognition, essential resources will degrade, knowledge will be lost, and the pace and reliability of discovery will suffer.

\item
\addcontentsline{toc}{subsection}{Evaluate scientific value}
{\bf Evaluate scientific value.}
Meaningful evaluation should determine whether AI-enabled ecosystems improve scientific and engineering practice, not merely production speed or volume. 
Relevant measures include the validity and usefulness of resulting scientific insights; the transparency and interpretability of computational processes; validation and uncertainty information; opportunities for human intervention; sustained support for essential people and infrastructure; and effects on collaboration and breadth of inquiry.
\end{titemize}

Near-term work should prioritize shared research assets, trust infrastructure, explicit human-AI teaming norms, and incentives for stewardship. 
Pilots should test user-experience designs, agent coordination mechanisms, and cross-institutional governance, while longitudinal studies should examine workforce effects, professional roles, and scientific value.
This synthesis reflects recurring workshop themes rather than a formal consensus process or systematic review. 
The priorities should be tested and refined across disciplines, institutions, and deployment contexts.

\section{Conclusion}
The 2026 workshop reinforced and extended a central message from the 2025 report: 
the future of scientific computing depends on ecosystems, not isolated tools. 
AI-enabled systems may accelerate parts of scientific work, but speed alone is not progress. 
The relevant question is whether these systems support valid, interpretable, and creative discovery.

Meeting that standard requires more than technical capability. 
Scientific computing communities need reusable software, auditable workflows, robust VVUQ and traceability, boundary-spanning teams, workforce pathways that develop critical judgment, and incentives that reward stewardship. 
These needs will persist as AI systems, computing architectures, and development practices change.

The path forward is socio-technical co-design in practice: people, software, data, infrastructure, institutions, incentives, and norms must evolve together through feedback, adaptation, and stewardship. 
Done well, these ecosystems can accelerate discovery and expand scientific inquiry while
preserving the rigor, transparency, accountability, and collaboration on which progress and public benefit depend. 

\phantomsection
\section*{Acknowledgments}
\addcontentsline{toc}{section}{Acknowledgments}

This workshop was partially supported by the U.S. Department of Energy (DOE) Office of Science Distinguished Scientist Fellows Program. We especially thank our DOE contacts: Hal Finkel and David Rabson,
DOE Office of Advanced Scientific Computing Research (ASCR).
We thank Joerg Gablonsky (The Boeing Company) for insightful contributions to workshop discussions on next-generation ecosystems in scientific computing, which helped shape the ideas conveyed in this
report.
We are grateful to Paul Messina for detailed feedback on the document; his suggestions improved the precision and perspective of the report. 
We thank Gail Piper for editing this manuscript and Madison Broeker for creating the cover page and diagrams in Figures 1 and 2.

\newpage
\clearpage
\phantomsection
\bibliographystyle{unsrtnat} 
\addcontentsline{toc}{section}{References}
\bibliography{bibs/workshop-refs}

\bigskip
\bigskip

\appendix

\newpage
\section{Workshop Description}
\label{sec:workshop-description}

The 2026 Workshop on {\em Next-Generation Ecosystems for Scientific Computing: Harnessing Community, Software, and AI for Cross-Disciplinary Team Science} engaged over 45 cross-disciplinary experts in Chicago, IL, during April 14--16, 2026, to chart a path toward more powerful, sustainable, and collaborative scientific software ecosystems.
The following description is reproduced from pre-workshop materials and retains the original future-tense wording.

\subsection*{Workshop Charge}

\begin{wrapfigure}{R}{0.37\textwidth}
\vspace{-0.4in}
\includegraphics[width=0.37\textwidth]{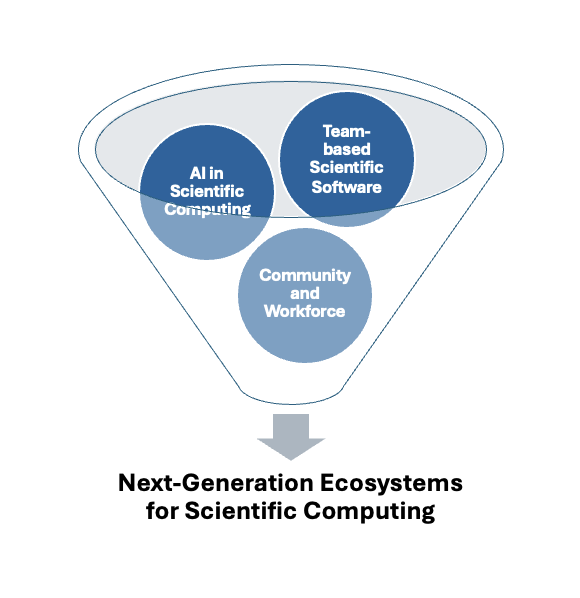}
\vspace{-0.4in}
\end{wrapfigure}
The high-performance computing (HPC) community has long driven scientific discovery at the limits of scale, complexity, and performance. Today, this leadership role is evolving rapidly as AI-enabled methods, heterogeneous architectures, and data-intensive workflows reshape how scientific computing is conducted. At the center of this transformation lies high-quality scientific software—the durable embodiment of domain expertise, computational methods, and collaborative practice that enables discovery to scale beyond individuals and institutions.

Motivated by urgent findings from recent community reports on scientific software development and AI for science, energy, and security, this workshop will convene cross-disciplinary experts spanning HPC, AI, computational science, applied mathematics, computer science, research software engineering, cognitive and social sciences, and community development. Participants will share, develop, and evaluate emerging strategies that address the challenges and opportunities shaping next-generation ecosystems for scientific computing.

\paragraph{Strategy.} This workshop represents Year 2 of a three-year series focused on strengthening team-based scientific software in an AI-driven future:

\begin{titemize}
\item Year 1 (2025): Identifying and understanding challenges, gaps, and opportunities
\item Year 2 (2026): Sharing, developing, and evaluating strategies and progress
\item Year 3 (2027): Coordinating ecosystem-wide advances that meld technical solutions and community-building
\end{titemize}
The 2026 workshop builds directly on insights from the 2025 workshop, as summarized in the report
\url{https://doi.org/10.48550/arXiv.2510.03413}, with an emphasis on leveraging synergies across AI, scientific software, and community initiatives to accelerate ecosystem-level impact.
The workshop series employs a co-design methodology, intentionally weaving together the following:

\begin{titemize}
\item Team-based scientific software development
\item AI-enabled scientific workflows and software infrastructure
\item Community, workforce, and ecosystem development
\end{titemize}
This integrated approach ensures that technical innovation and community evolution advance together, enabling broad, sustainable, and impactful scientific computing ecosystems.

\paragraph{Areas of emphasis.} Our goal is to assess, strengthen, and transform scientific software ecosystems, informed by state-of-the-art team science, to meet the needs of next-generation research in scientific computing, while responsibly advancing emerging AI technologies. Technical and community perspectives will be integrated throughout discussions on the following themes:

\begin{titemize}
\item {\bf Software and next-generation science}
\begin{titemize}
\item Examining how new scientific frontiers, AI-augmented workflows, and heterogeneous computing platforms demand new approaches to software and workforce development
\item Embedding community-driven practices to ensure that technical solutions reflect broad expertise, use cases, and stakeholder needs
\end{titemize}

\item {\bf AI-driven software ecosystems for scientific computing}

\begin{titemize}
\item Mapping pathways toward robust, interoperable, and scalable software ecosystems that enable AI-driven discovery in HPC and data-intensive environments
\item Advancing AI tools, frameworks, and infrastructure for scientific computing that address current gaps while supporting openness, reproducibility, and broad participation
\end{titemize}

\item {\bf Team-based software and cross-disciplinary research}

\begin{titemize}
\item Identifying evolving roles, career paths, and best practices for collaborative scientific software teams operating at the intersection of AI and computational science
\item Emphasizing community co-design, where software evolves through continuous dialogue among scientists, developers, users, and maintainers
\end{titemize}

\item {\bf AI for scientific software productivity and sustainability}

\begin{titemize}
\item Exploring how AI-assisted development, testing, performance tuning, documentation, and maintenance can enhance productivity and long-term software sustainability
\item Embedding feedback loops that align AI-enabled tools with real-world scientific computing workflows and community needs
\end{titemize}

\item {\bf Community and workforce development}

\begin{titemize}
\item Accelerating strategies to cultivate next-generation R\&D teams equipped to operate in AI-rich scientific computing environments
\item Fostering collaborative cultures that bridge technical excellence with broad, community-centered ecosystem growth
\end{titemize}
\end{titemize}

\subsection*{Workshop Objectives} 

The workshop aims to shape a forward-looking, actionable vision for the future of team-based software in scientific computing, grounded in both emerging AI capabilities and the realities of cross-disciplinary collaboration. Key objectives include the following:

\begin{titemize}

\item
{\bf Advancing scientific software practices}: Building shared understanding of emerging methodologies, tools, and norms for team-based scientific software development, integrating technical rigor with community sustainability.

\item {\bf Creating strategies for excellence}: Curating and refining resources, frameworks, and exemplars that support excellence, effectiveness, and resilience in scientific software collaborations.

\item {\bf Overcoming challenges and creating opportunities}: Identifying strategies to address barriers faced by scientific software teams while enabling innovation aligned with emerging research and AI-driven needs.

\item {\bf Envisioning the future}: Articulating a shared, forward-looking vision for next-generation scientific software ecosystems that harmonizes technical innovation with community dynamics.

\item {\bf Fostering community development}: Defining concrete actions to strengthen and broaden the workforce, while fostering durable communities prepared to meet urgent challenges in HPC and AI-enabled scientific computing.

\end{titemize}

\noindent
By intentionally integrating technical innovation with community-centered design, this workshop seeks to shape a future in which thriving, cross-disciplinary ecosystems drive the next wave of scientific discovery through advanced computing. In this future, high-quality scientific software—co-designed, AI-enabled, and sustainably maintained—serves as the keystone of enduring collaboration and scientific progress.

\subsection*{Workshop Outcomes}

Participants will share and examine perspectives, experiences, and emerging practices related to team-based scientific software in an AI-enabled future, with the goal of assessing progress, identifying remaining gaps, and prioritizing areas for continued attention. Through discussion and co-design activities, the workshop will emphasize high-impact focus areas, spanning technical, organizational, and community dimensions.

Insights emerging from the workshop will be synthesized in a post-workshop report, extending the findings of the 2025 workshop. This report will contribute to a growing body of community knowledge intended to inform ongoing and future efforts in scientific software development, workforce advancement, and ecosystem coordination. It also  may help shape evolving perspectives, policies, and practices across the scientific computing community.

By intentionally interweaving community considerations with technical discussions throughout the workshop series, this three-year effort aims to expand and strengthen the scientific software community, foster cross-disciplinary collaboration, and help accelerate next-generation scientific discovery in an increasingly AI-driven world.

\newpage
\section{Workshop Participants}
\label{sec:workshop-participants}

\subsection*{Organizing Committee}

\begin{titemize}

\item Lois Curfman McInnes, Argonne National Laboratory (ANL), Senior Computational Scientist,
Mathematics and Computer Science (MCS) Division

\item Dorian Arnold, Emory University, Associate Professor, Department of Math and Computer Science

\item Prasanna Balaprakash, PrimaLabs, Co-founder and CEO

\item Mike Bernhardt, Team Libra, Founder and Chief Strategy Officer; 
former Director of Communication for the DOE Exascale Computing Project

\item Franck Cappello, ANL, Senior Computer Scientist, MCS Division

\item Beth Cerny, ANL, Head of Communications for the Argonne Leadership Computing Facility (ALCF)

\item Anshu Dubey, RIKEN Center for Computational Science

\item Denice Ward Hood, University of Illinois Urbana-Champaign, Associate Professor, Department of
Education Policy, Organization \& Leadership

\item Mary Ann Leung, Sustainable Horizons Institute, Founder and President

\item Olivia B. Newton, University of Montana, Faculty, Department of Management Information Systems

\item Keita Teranishi, Oak Ridge National Laboratory (ORNL), Group Leader of Programming Systems, Computer Science and Mathematics (CSM) Division

\item Stefan M.\ Wild, Lawrence Berkeley National Laboratory (LBNL), Director, Applied Mathematics and Computational Research Division
\end{titemize}

\subsection*{Attendees}

\begin{titemize}
\item Gabrielle Allen, University of Wyoming, Professor, School of Computing

\item Richard Arthur, GE Aerospace, Senior Principal Engineer, Computational Methods Research

\item Alexandra Ballow, Montana State University, Doctoral Student, Mathematical Sciences

\item Tony Baylis, Lawrence Livermore National Laboratory (LLNL), Senior Manager

\item  David E. Bernholdt, ORNL, Distinguished R\&D Staff Member, CSM Division

\item Daniel Bielich, Synopsys, Research and Development Engineer

\item Johanna Cohoon, LBNL, User Experience Researcher, Scientific Data Division

\item Jeremy Crampton, AAAS Fellow, U.S. Department of Energy, Advanced Scientific Computing Research

\item Deborah DiazGranados, Virginia Commonwealth University, Industrial/Organizational Psychologist and
Associate Professor, School of Medicine

\item Nichole Etienne, Emory University, Doctoral Student, Department of Math and Computer Science
 
\item Charles Ferenbaugh, Los Alamos National Laboratory (LANL), R\&D Scientist, Applied Computer Science

\item Stephen M. Fiore, University of Central Florida, Professor, Director of Cognitive Sciences Laboratory

\item Joerg Gablonsky, The Boeing Company, Technical Fellow, Technical Lead Numerical Optimization, Chair of Enterprise HPC Council
 
\item Roscoe Giles, Boston University, Professor, Electrical and Computer Engineering, also Computing and Data Science

\item Diego G\'{o}mez-Zar\'{a}, University of Notre Dame, Assistant Professor, Computer Science \& Engineering

\item Thomas Herault, INRIA, Senior Researcher
 
\item Tanzima Islam, Texas State University, Assistant Professor, Department of Computer Science

\item Stephen Jacobsohn, Google, Principal Enterprise Architect

\item Meifeng Lin, Brookhaven National Laboratory, Chair of Computational Science Department

\item Charles Lively, LBNL, Science Engagement Engineer and HPC Consultant, NERSC
 
\item Vanessa López-Marrero, Stony Brook University, Computational Research Scientist, Institute for Advanced Computational Science

\item Satoshi Matsuoka, RIKEN Center for Computational Science, Director 

\item Staša Milojević, Indiana University, Professor, Informatics

\item Daniel Nichols, LLNL, Sidney Fernbach Postdoctoral Fellow

\item Christopher Oehmen, Pacific Northwest National Laboratory, Senior Research Scientist and Group Leader, Biological Sciences Division

\item Santiago Ospina Tabares, University of Illinois Urbana-Champaign, Doctoral Student, Digital Environments for Learning, Teaching, \& Agency

\item Michael Papka, ANL, Deputy Associate Laboratory Director, Computing, Environment and Life Sciences;
Director of ALCF; University of Illinois Chicago, Professor, Computer Science

\item Irene Qualters, LANL, Associate Laboratory Director Emerita, Simulation and Computation

\item Katherine Riley, ANL, Director of Science for ALCF

\item Damian Rouson, LBNL, Group Lead for Computer Languages and System Software, AMCR Division

\item Sudip Seal, ORNL, Group Lead for Systems and Decision Sciences, CSM Division

\item Brittany Segundo, National Academies of Sciences, Engineering, and Medicine, Portfolio Lead, Mathematical \& Statistical Sciences

\item John Shalf, LBNL, Department Head for Computer Science

\item Andrew Siegel, ANL, Senior Computational Scientist, MCS Division and Nuclear Engineering Division

\item Valerie Taylor, ANL, Director of the Mathematics and Computer Science (MCS) Division 

\item Jim Willenbring, SNL, Senior Member of R\&D Technical Staff, Center for Computing Research

\item Lou Woodley, Center for Scientific Collaboration and Community Engagement, Founder and Director

\end{titemize}

\subsection*{DOE Contacts}

\begin{titemize}
\item Hal Finkel, U.S. Department of Energy (DOE), Office of Advanced Scientific Computing Research, Associate Director

\item David Rabson, DOE, Office of Advanced Scientific Computing Research, Physical Scientist

\end{titemize}

\newpage
\section{Workshop Agenda}
\label{sec:workshop-agenda}

\includegraphics[scale=0.82,page=1,trim={0.8cm 0cm 2.0cm 1.5cm},clip]{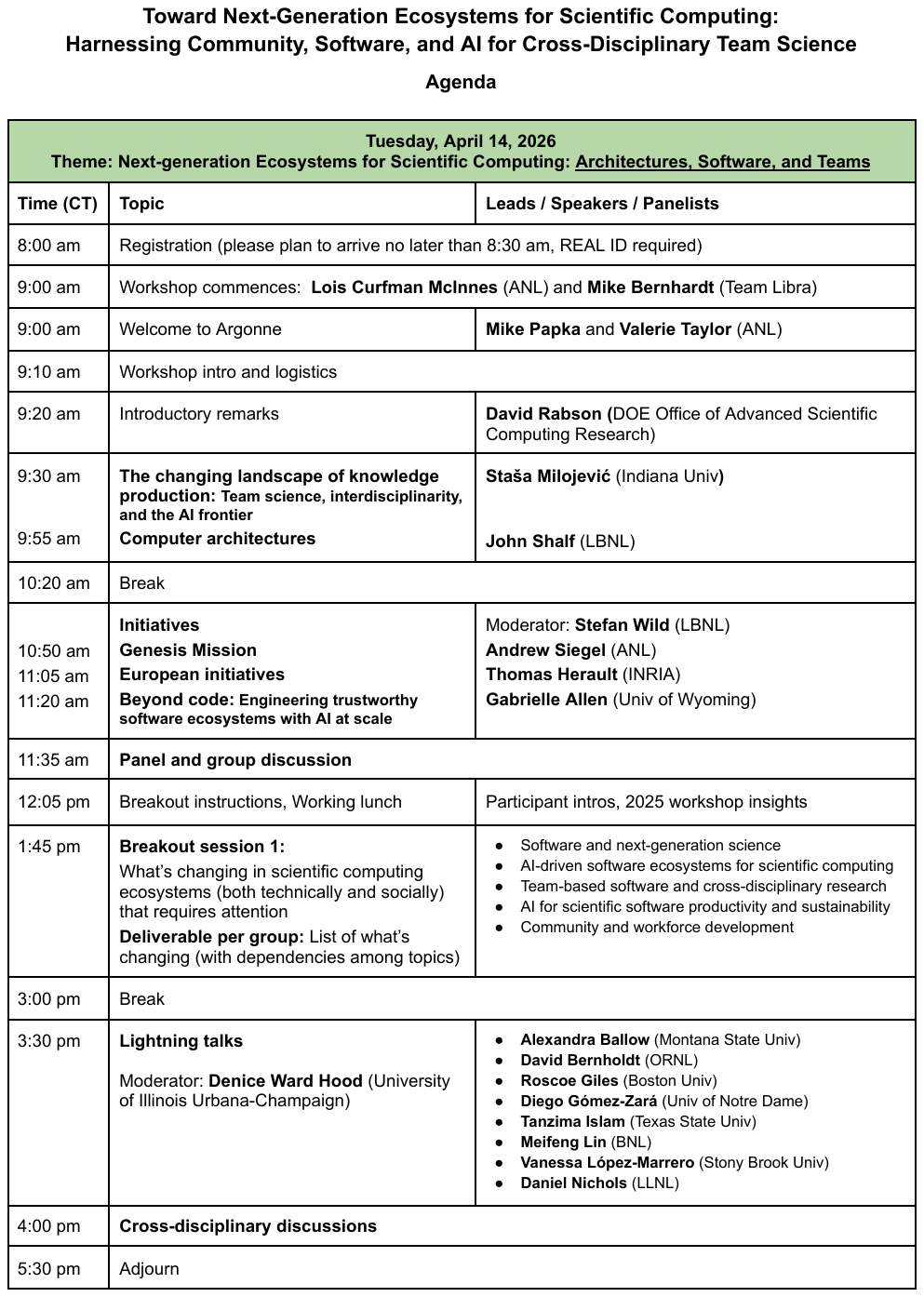}
\newpage

\includegraphics[scale=0.82,page=2,trim={1.4cm 0cm 2.0cm 1.5cm},clip]{figures/Agenda.Workshop.Next-GenEcosystems.2026.04.pdf}
\newpage

\includegraphics[scale=0.82,page=3,trim={1.4cm 0cm 2.0cm 1.5cm},clip]{figures/Agenda.Workshop.Next-GenEcosystems.2026.04.pdf}

\includegraphics[scale=0.82,page=4,trim={1.4cm 0cm 2.0cm 1.5cm},clip]{figures/Agenda.Workshop.Next-GenEcosystems.2026.04.pdf}

\includepdf[pages=1]{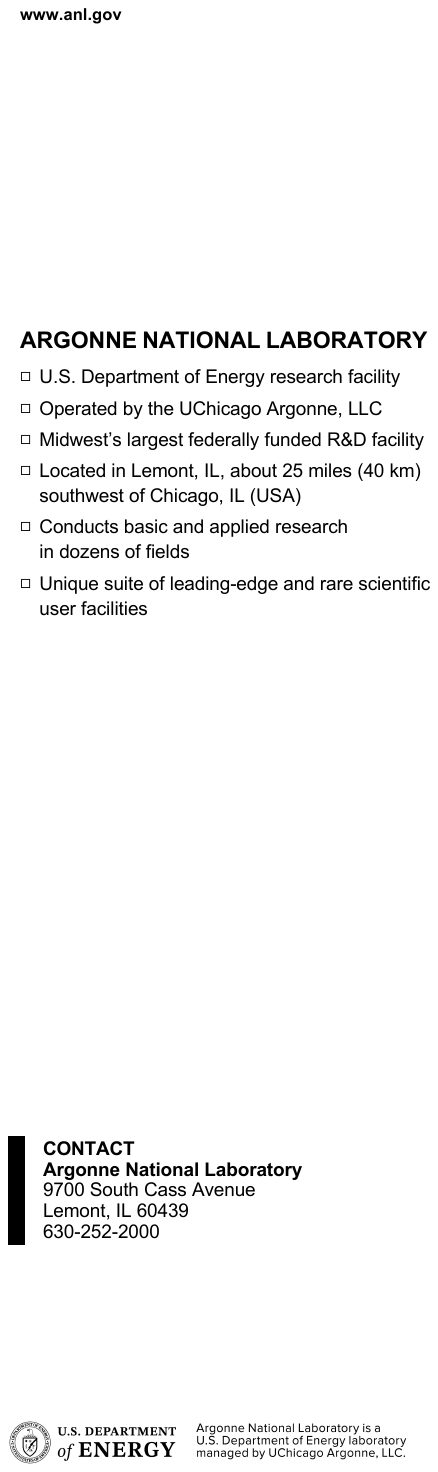}

\end{document}